\documentclass[conference]{IEEEtran}
\IEEEoverridecommandlockouts
\usepackage{cite}
\usepackage{amsmath,amssymb,amsfonts}
\usepackage{algorithmic}
\usepackage{graphicx}
\usepackage{textcomp}
\usepackage[T1]{fontenc}
\usepackage{verbatimbox}
\usepackage{float}
\usepackage{placeins}
\usepackage{listings}
\usepackage{textcomp}
\usepackage{xcolor}
\usepackage{caption}
\usepackage{placeins}
\usepackage{nameref}
\usepackage{varioref}
\usepackage{hyperref}
\usepackage{cleveref}
\usepackage{lipsum}
\usepackage[utf8]{inputenc}
\usepackage{fancyhdr}
\usepackage{mathptmx}
\newcounter{counter}[section]

\usepackage[per-mode=symbol]{siunitx}
\usepackage{bm}

\definecolor{codegreen}{rgb}{0,0.6,0}
\definecolor{codegray}{rgb}{0.5,0.5,0.5}
\definecolor{codepurple}{rgb}{0.58,0,0.82}
\definecolor{backcolour}{rgb}{0.95,0.95,0.92}

\lstdefinestyle{mystyle}{
    backgroundcolor=\color{backcolour},   
    commentstyle=\color{codegreen},
    keywordstyle=\color{magenta},
    numberstyle=\tiny\color{codegray},
    stringstyle=\color{codepurple},
    basicstyle=\ttfamily\footnotesize,
    breakatwhitespace=false,         
    breaklines=true,                 
    captionpos=b,                    
    keepspaces=true,                 
    numbers=left,                    
    numbersep=5pt,                  
    showspaces=false,                
    showstringspaces=false,
    showtabs=false,                  
    tabsize=2
}

\usepackage{xcolor}
\def\BibTeX{{\rm B\kern-.05em{\sc i\kern-.025em b}\kern-.08em
    T\kern-.1667em\lower.7ex\hbox{E}\kern-.125emX}}
\begin{document}
\title{An Analysis Of Bugs In Persistent Memory Application\\
}

\author{\IEEEauthorblockN{Jahid Hasan}
\IEEEauthorblockA{\textit{Department of Computer Science} \\
\textit{Iowa State University}\\
}
}

\maketitle
\thispagestyle{plain}
\pagestyle{plain}
\begin{abstract}
Over the years of challenges on detecting the crash consistency of non-volatile persistent memory (PM) bugs and developing new tools to identify those bugs are quite stretching due to its inconsistent behavior on the file or storage systems.
In this paper, we evaluated an open-sourced automatic bug detector tool (i.e. AGAMOTTO) to test NVM level hashing PM application to identify performance and correctness PM bugs in the persistent (main) memory. Furthermore, our faithful validation tool able to discovered 65 new NVM level hashing bugs on PMDK library and it outperformed the number of bugs (i.e. 40 bugs) that WITCHER framework was able to identified. Finally, we will propose a Deep-Q Learning search heuristic algorithm over the PM-Aware search algorithm in the state selection process to improve the searching strategy efficiently.
\end{abstract}

\section{Introduction}\label{section:lab}
Recently, Persistent Memory (PM) caught the attention of many researchers in terms of comparing the performances with existing non-volatile memory (NVM) technologies. Intel Optane DC introduces this PM, which is accessed with byte-addressable non-volatile memory and it tends to be much faster and cheaper than traditional Dynamic Random Access Memory (DRAM). With lower latency rate and higher bandwidth, it improves the crash recovery immensely. Even with the advantages, it does introduce persistency memory bugs on the program due to its improper use cases of memory store, flush/fences which leads to crash consistency on memory devices, that brings a huge challenges to analyze those bugs either in a static or dynamic way to exploit the program code efficiently. Some of the existing PM applications available to perform significant analysis to diagnose those crash consistency bugs are pmemcheck\cite{pmdk}, PMFS\cite{pmfs}, PMDK\cite{pmem} library, respectively.

\smallskip
Based on previous study and existing solutions such as Intel Yat\cite{yat}, pmemcheck\cite{pmdk}, PMDK
(Persistent Memory Development Kit)\cite{pmem}, WITCHER\cite{witch}, PMTest\cite{pmtest}, XFDetector\cite{xfd} still left with some limitations on their own. Its obvious that designing and implementing such a sophisticated model to detect the root cause of such failure and at the same time correcting those PM bugs are quite complicated for PM developers to trigger it out properly. There are certain parameters that can be considered on application specific bugs based on its correctness and performance behaviour that usually annotated by those debugger tools represents as a false positives or false negatives. We evaluated our NVM based level hashing PM application on AGAMOTTO\cite{neal} to validate the testing performances and to compare our testing results with WITCHER\cite{witch} testing framework results. Basically, we borrowed the level hashing benchmark from WITCHER artifact and to make it compatible and executable on AGAMOTTO we did minor changes on their source code for our experiment which we will discuss later section. Based on our testing experiment, we able to detect 65 new PM bugs in the PMDK library which outperforms WITCHER testing reports. The bug reports that generated by our faithful detector was 60 unpersisted write bugs (correctness), 2 flush (performance), 3 fence (performance) bugs along with real-world systems. On the other hand, our key performance evaluation with WITCHER level hashing benchmark, which was actually able to detect 40 new bugs while majority counted bugs unpersisted correctness bugs (i.e. 17).

\smallskip

The primary focus of this paper is to study depth of PM bugs in different PM (main) software applications, then test the level hashing PM application on AGAMOTTO to analysis the behavior of bugs that occurred on the NVM programs, and finally pointing out a novel approach of the state representation where we will introduce Deep-Q learning algorithm to make the priori knowledge search more robust and faster than the PM-Aware search algorithm. 

We make the following contribution of this work as follows:
\begin{itemize}
    \item We conduct a thorough study on different persistent (main) memory bugs on PMDK library and present their detailed study.
    \item We perform our implementation on AGAMOTTO to test the NVM level hashing data structure to detect and maintain the application specific bugs automatically.
    \item Our evaluation found 65 new bugs in a single NVM level hashing application using our validation tool compared with the WITCHER testing results. Also the resulting report has no false positives bug.
    \item Finally, we propose a Deep-Q learning algorithm over PM-Aware search algorithm for better state representation and selection to update the heuristic search tree in a robust manner to make decision.
\end{itemize}

Finally, the rest of this paper is organized as follows. Section~\ref{section:back} will provide more background study of PM programming bugs, PM libraries, mode operations, level hashing, proposed algorithm design. Section~\ref{section:imp} describe the key implementational process. Then, Section~\ref{section:eval} present the experimental results that obtained by our faithful automatic debugger, provide with insightful details of each bug threads pattern to detect correct NVM level hashing bugs. Then, Section~\ref{section:con} conclude with a few outcome of this project and with possible future research opportunities in this specific area of bug detection.

\section{Background And Motivation}\label{section:back}
In this section, we will first introduce the persistent (main) memory systems and its persistency bugs, mode operations workflow. We will also discuss level hashing data structure and our proposed heuristic searching algorithm over PM-Aware search algorithm.

\subsection{Persistent Memory (PM)}
To begin with the background study of non-volatile memory (NVM) technologies in general, we first introduced us with the Intel Optane DC memory for increasing our storage capacity with lower latency rate in a persistent manner. That means even if a crash or failure occurs in the system, this optane DC memory can have data recovery state and that makes it more unique and powerful than any other existing traditional DRAM. As for our study, we will scrutinize the PMDK library in essence, as its widely maintained by Intel and also it used by many researchers to use their library packages which build on DAX (Direct Access) to analyze the memory-mapped NVM PM programs. PM programs are consists of few crash-consistency instructions such as $store$, Cache-Line Write-Back $(CLWB)$, $Flush/Fence$ \cite{fast}. Basically, here a store acts as a program on persistent memory address to manipulate its location address, while the $CLWB$ instructions writes back the cache line with a specific location back to the memory controller by using the $sfence$ to complete the cache hierarchy on the dirty memory locations. These operations or instructions ensures that the persistent memory addresses are has been properly persisted or not by enforcing the $fence$ instructions improperly.

In PM software based application, its always a challenging tasks to provide a much efficient and scalable consistent memory support even a persistency bugs found in our applications. As a researcher perspective, the behavior of those persistency bugs in our source code and detect it by using an handy tool to validate and reduce the threat of those persistency memory bugs are always quite challenging for the developers. In this project, we will use our faithful bug detector tool- AGAMOTTO, which is a bug detector tool to automate the process and validate the use case in practice to correct and error-prone the crash-consistency bugs in our PM application. As persistence memory are categorized in two distinct categories by many researchers to evaluate the PM bugs, which also known as PM correctness and performance bugs.

Based on above classification, we can illustrate that the persistence correctness bugs which violates the crash consistency behavior (such as: ordering or atomicity PM bugs) can leads to a inconsistent action of the NVM systems and at the same time it will fail to recover the data even after a power crash occurred. On the other hand, the performance bugs are unnecessary persistency program bugs which degrade the performance of the applications. These type of performance bugs are mostly unpersisted or flush/fence type of performance bugs.

\subsection{PM Operations Mode}
\begin{figure}[htbp]
\centering
    \includegraphics[width=0.52\textwidth]{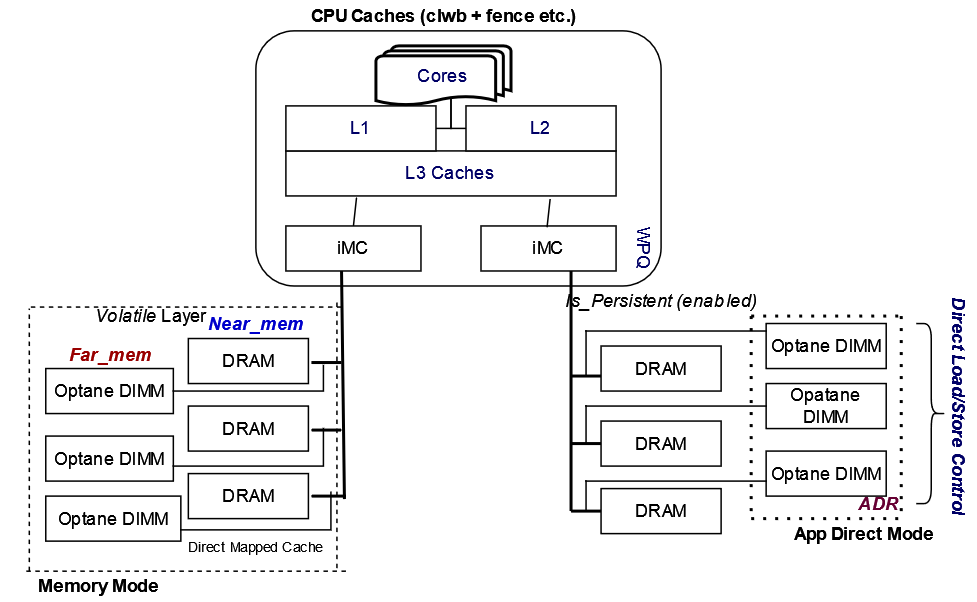}
    \caption{Optane PM Mode Operations Workflow.}
    \label{fig:pm}
\end{figure}

In recent years, Intel Optane Dual In-Line Memory Module (DIMM) became $2^{nd}$-Gen scalable processor with lowest memory latency, higher bandwidth of byte-addressable memory~\cite{key}. In general, the PM writes are done on the caches in a volatile write back CPU caches while it does not flushed to PM at once to make it persisted. In such case, the ordering of cache flushes does not provide any gurantee on PM write back cache properly. Although, the PM holds the data being persisted and consistent, so it can recover its data even after a power failure or crash consistency occurred in the memory storage systems. Thus, to ensure an efficient or the consistent behavior on PM systems, developer must place the appropriate Cache-Line Write-Back, flushes (e.g., $CLWB$), and fences (e.g., $sfence$) right after a PM writes. In figure~\ref{fig:pm}, it shows that unlike the conventional DRAM DIMMs, the Intel Optane DIMM installed on the memory bus, and its communicated via the Integrated Memory Controller (iMC) block. Based on the mode operations workflow, it displays that each iMC combines with total of six channels and two different operating modes as Memory Mode and App Direct Mode \cite{intel}. At the beginning, we can see that the CPU caches consists of $CLWB +$ $Fence$ on top of the CPU cores. Each CPU cores combined with $L1$ and $L2$ caches and in the below layer it has one shared $L3$ caches connected with iMC for each two PM operating modes. To ensure the persistency, the communication block of iMC try to maintain the Read and Write Pending Queues (RPQs and WPQs) for each of the Intel Optane DIMMs and which also includes the Asynchronous DRAM Refresh (ADR) that associated with the iMC block. This ADR will make sure that data reaches evenly as it does not contain any processor caches and can store it even there is a power crash occurred as iMC will flush the updates beforehand.
\smallskip

The two Intel Optane operating modes are described as follows:
\subsubsection{Memory Mode}
On this PM memory mode operation from the figure, we can see that it comprises with Optane DIMM which is a far dense slower memory than the traditional DRAM which is near memory location direct mapped with cache for the Non-Volatile (NVDIMM). The cache block size is 64 B, and also the cache is handled opaquely by the CPU's main memory controller. Memory mode uses Intel Optane DIMMs to expand its main memory size without persistent as its located in the volatile layer. That means data is lost if a power failure event occurs.

\smallskip

\subsubsection{App Direct Mode}
On the other hand, in the app direct mode it checked $Is\_Persisted$ is enabled or not, as it separate the traditional DRAM and Optane DIMMs. We can also observe that, the DRAM does not contain any cache in persistent memory. A file system or other management layer utilizes persistent data by allocating (i.e. direct memory load/store control operations), naming, and accessing it at lowest latency. To make the data or storage stack persistent on ADR domain, its required to use app direct mode operation to enable the persistent memory-aware file systems. 

\smallskip

Finally, in both PM operating modes, the Intel Optane memory can be interleaved across each channels and DIMMs as well.

\subsection{Level Hashing Data Structure}
To detect persistent memory bugs in NVM based program, we used level hashing based key-value pair data structure to detect bugs using our validation tool. Generally, level hashing~\cite{level} enables the consistency of log-free write transactions to maintain the key-value pairs. In other words, it is a write-optimized, high-performance hashing index mode with a minimum consistency guarantee and cost-effective resizing for persistent memory. We used the level hashing benchmark from the original WITCHER artifact to perform our testing. Based on our faithful debugger, we can observe that the reproduces of PM bugs are mostly found due to the incorrect ordering, atomicity between two metadata and hashing. A crash in the center of a rehashing execution in level hashing, may result inconsistent with both retrieved key-value pair metadata lost or the redundancy issue.

\begin{figure}[htbp]
\centering
    \includegraphics[width=0.45\textwidth]{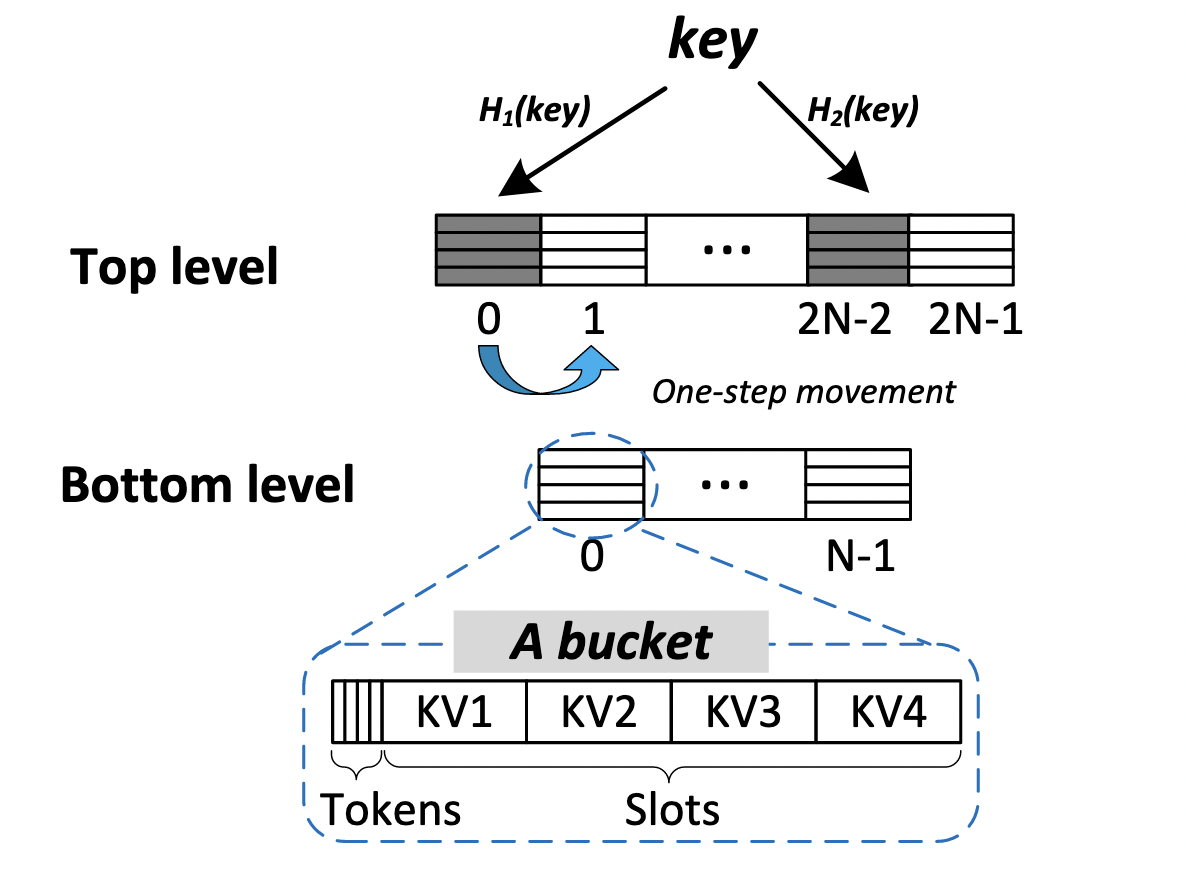}
    \caption{Overview of Level Hashing Structure \cite{lvl}.}
    \label{fig:level}
\end{figure}

Followed by figure~\ref{fig:level}, we can see that a level hashing structure consists of two levels to provide low-overhead crash-consistency, efficient write log and resizing of the data structure as well. At first, we can see at the bottom level which consists of 4 bucket slots, such as $(KV1, KV2, KV3, KV4)$ and their corresponding tokens, size of one bit for each slot for the crash-consistency behavior. From this data structure representation, it mostly has two independent hash functions to store the value on each buckets consistently, while these buckets become occupied then the level hashing enact one-step movement to create an empty slot to insert an item from the buckets. While at top level, it comprised with two keys such as $H_1, H_2$ keys for two individual buckets to perform such insertion and storing movement from one block to another. Basically the one-step movement helps level hashing to improving its maximal load factor prior to perform resizing \cite{level}. To perform resizing operation for maximum efficiency, the top level creates as many as buckets to store the items while those items are re-utilize again without a rehashing operation. To achieve maximum concurrent queries for each slot-grained block they perform duplicate (violates the update and deletion rules that are not being modified) and missing (missing subsequent inserted items due to concurrent queries that move from one thread to another) correctness problems. 
\smallskip

For better understanding, below we showed a listing 1 of non-atomic initialize type of inconsistent bug that observed on WITCHER framework, where we can see that most of the cases it was inconsistent because of its improper use of $NULL$ checks on the $init\_nvmm\_pool()$ function which lead to segmentation fault on it while its creating PMDK API pool and improper $root\_obj$ allocation on level hashing key-value pairs. Which is specifically illustrated as data structure related bugs on level hashing program.

\begin{lstlisting}[language=Octave]
  # No "NULL" check added on "init_nvmm_pool" to return a value
    root_obj *root_obj = init_nvmm_pool(path, size, layout_name, &is_created);
    if (is_created == 0) {
      printf("Reading from an existing level hash.\n");
      return root_obj->hash_ptr;
      
      
      
\end{lstlisting}
\captionof{lstlisting}{ Non-atomic initialization of NULL bugs adapted from WITCHER $Level\_hashing.c$ Issue $\#1.$}

\subsection{Deep-Q Learning Algorithm}
\begin{figure*}[!htbp]
    \centering
    \includegraphics[width=1\textwidth]{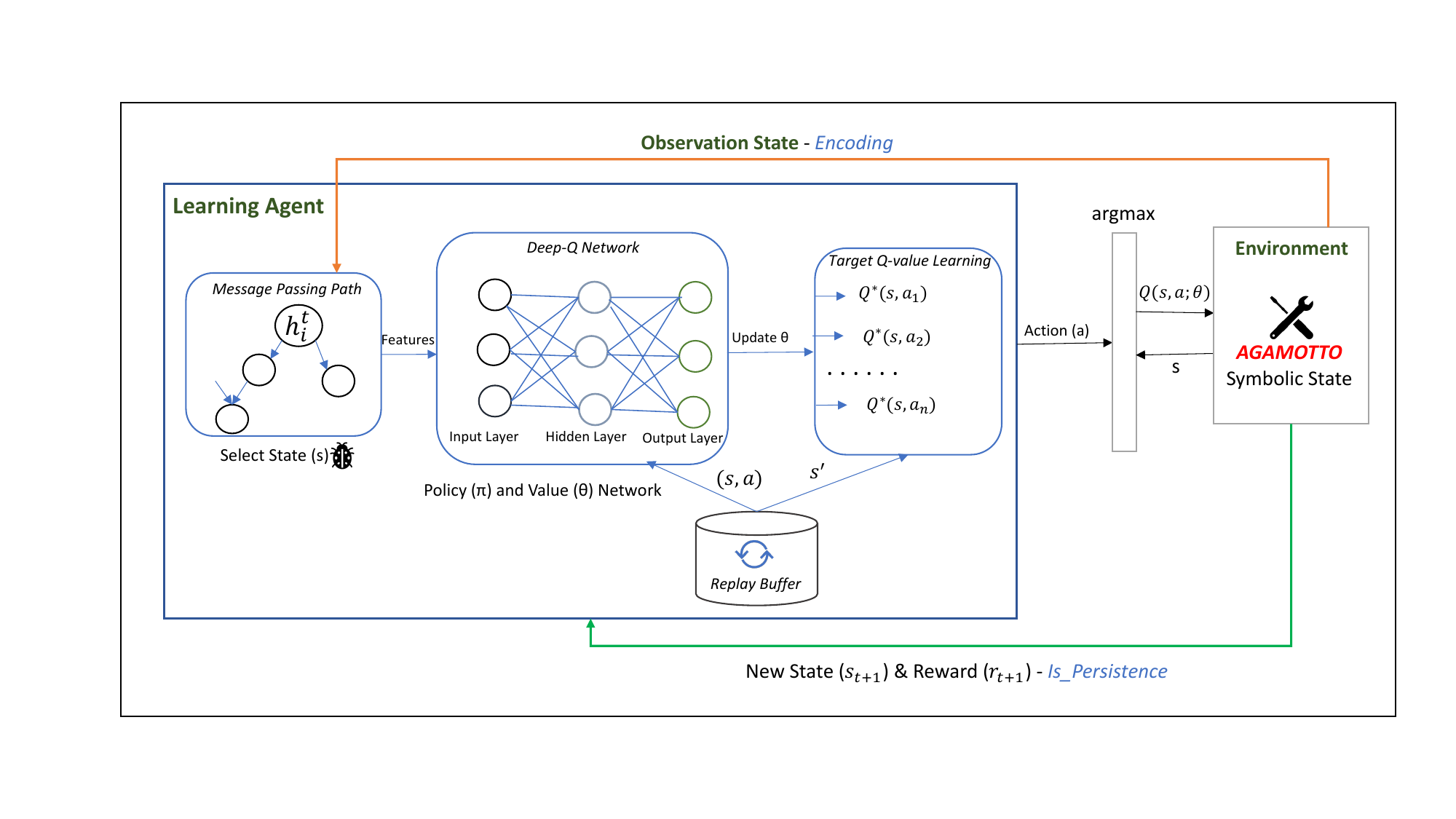}
    \caption{Proposed RL Based Deep-Q Learning Model Workflow.}
    \label{fig:model}
\end{figure*}

In this section, we will propose our idea of improving existing PM-Aware search algorithm based on our study from Ayudante~\cite{ayun} paper, which considered to be the first paper which brings the idea of using Reinforcement Learning (RL) to improve the search strategy. Here, we will emphasize on Deep-Q learning and the message passing path searching strategy. We will try to update the AGAMOTTO symbolic execution path engine based RL framework to further analyze and prioritize the computation based on transfer learning knowledge from one state to another by an state-representation of neural network in our tentative proposed model. Figure~\ref{fig:model} shows our proposed method of designing RL based Deep-Q learning model workflow that generate PM code from trained network model and validate it from our environment. The key components of our model consists of a learning agent and an environment. The process of our model training, the learning agents try to perform certain actions to the environment based on the state and reward it received as well as the observation from state itself. We will discuss it in detailed below:
\smallskip 

\textbf{Message Passing Path Algorithm:} First of all, we begin with our user specified message passing path as a select state representation $(s)$, which computes all the node features and the connected nodes embedding from edges. In Graph Neural Network (GNN)~\cite{gnn} this message passing path algorithm are used to propagate the state information each of the nodes of graph in an iterative manner. The key idea of each node is to receive corresponding neighbors information or data even aggregating the hidden states of each node itself. From figure, we are selecting our state representation $(s)$ on message passing path based on one step forward of each message at a time stamp $(t)$, while the hidden state vector of all node features are $h_i$ respectively from their neighbors node indicating the function of $h^{t}_{i}$. 
\smallskip

\textbf{Deep-Q Network Learning:} In this stage, we are passing all of our selected state of features to our Deep-Q network (DQN)~\cite{dqn} model free network which consists of input, hidden, and output layers to learn the approximate Q-value of each actions and updates based on our policy $(\pi)$ and value $(\theta)$ without not being generalized  with other RL models. Here, the policy function indicate our learning agent behavior over model that associated with from state to an action, while the value function quantify the expected reward from a state $s$ for $\pi$ value helps the model to do the prediction what the symbolic state does in the environment with an good approximation rate. When we update $\theta$ periodically from our DQN to the target Q-value learning, we will get an optimal $Q^*$ function for different actions $(a_1, a_2,...,a_n)$ on selected state $(s)$ pair. We can represent it as follows using Bellman equations to update the Q-values:
$$Q^*(s,a) = (1-\alpha) Q(s,a) + \alpha (r(s,a) + \gamma . \max_{\alpha} Q^*(s_{t+1}, a^\prime)) $$

Where, $Q^*(s,a)$ is the optimal function from select state $(s)$ to an action $(a)$ to find an optimal policy from our equation. The decay rate is $\alpha \in \{0,1\}$ and the discount factor is $\gamma \in  \{0,1\}$~\cite{ayun}. Then we will obtain reward value by selecting state to an action such that the updated new optimal function value is $Q^*(s_{t+1}, a^\prime)$. we also used the replay buffer to synchronize the updates properly as Q-value table become larger and cause some issue on learning policy value from high-dimensional selected state and actions~\cite{dqn}. So, this replay buffer will work as an controller estimator for Q-value on the network space by storing previous state outputs.
\smallskip

Once our model training finished, it then perform some action to argmax before pass to the environment. The key idea here is to find the maximum value from higher nodes from a given input functions. It uses the idea of leaf embedding. It then calculate the expected reward function to pass on the environment state and also return a state value back to argmax. The symbolic state on environment then propagated back to our message passing path to encode the observation state. The encoding can be perform by using LSTM encoder-decoder to perform better on our convolution neural network. On the other hand, we back propagate our updated new state and reward back to the learning agent to verify either the result is persistent or not.
\smallskip

Finally, the proposed model may perform better than the way Ayudante~\cite{ayun} proposed their work. Although, our model is quite ambitious to check PM bugs and provide refine reports over our faithful symbolic state tool. As Ayundante source code are not publicly available, so we did not able to perform any test to validate their claim/results, but this can be an interesting future work to perform. Overall, the model we have been proposed based on our study is quite compact and feasible to apply into Ayudante framework and integrate with AGAMOTTO PM-Aware algorithm. If this works, then we might able to detect PM program bugs quite efficiently and accurately than other existing work that have been published so far.

\FloatBarrier

\section{Implementation}\label{section:imp}
We set-up and performed all the testing using AGAMOTTO open-sourced artifact from GitHub, which built on top of KLEE\cite{klee}. As this artifact comprises with over $\sim 3000$ LOC (lines of code) of C++ and expanding the KLEE which is around $\sim 4000$ LOC of C++ that has been written. We also used WLLVM (Whole-Library Low-Level Virtual Machine) to generate LLVM specific byte-code for the NVM level hashing program. Based on artifact documentations, we first build the virtual machine (VM) on our VM server to reproduce the AGAMOTTO original reported results on its five different PM applications to observe the results on different runs, mostly run for about $2\sim5$ minutes to cover minimum number of exploration path to detect the bugs, although the recommended max-time period was for 60 minutes for each test run. Based on our compilation setup, we faced execution issue with memcached-pmem-prebuilt, pmdk-prebuilt which stucked at some point and failed to generate the exact results while it run for 60 minutes. While other three systems run properly and generate same results as the original paper stated. But, for this piece of work, we borrowed the WITCHER level hashing program to experiment on AGAMOTTO and compare these two frameworks testing results on total number of bugs that it can calculated. In our experiments, AGAMOTTO artifact works much better than the WITCHER framework. Apart from that, the WITCHER framework was modified into around 3,600 lines of C++ code while level hashing lines of C code was $\sim 600$.

\section{Performance Evaluation}\label{section:eval}
This section presents the detailed evaluation of our experimental results and listing a few pmem bugs reports based on bug classifications in details. 
\subsection{Evaluation Setup}
Our experiments run on 11th Gen Intel(R) Core(TM) i7-11800H@ 2.30GHz, NVIDIA
GeForce RTX 3050 Ti GPUs with 32 GB RAM. We evaluated our faithful debugger tool on dedicated \textit{Ubuntu 20.04.1 LTS} server, where we implemented and tested the open-source artifact of AGAMOTTO to perform certain testing on specific persistent level hashing benchmark program from WITCHER GitHub repository on a headless \textit{Ubuntu 18.04.5 LTS} server. We also installed GNU gdb debugger to test level hashing bug threads manually on $Ubuntu 8.1.1-0ubuntu1$ to locate the PM bugs static stack frame bugs that occurred during the test run.

\subsection{Experimental Results Analysis}
In this section, we will evaluate our testing results on AGAMOTTO as it has consistent and better accuracy to detect different PM applications and libraries such as Intel PMDK library. In addition to PMDK library, we tested NVM level hashing program borrowed from WITCHER framework on our faithful debugger tool. The total testing run was executed for $60$ minutes and our bug detector tool able to explored all the paths within $5$ minutes and able to confirm 65 new PM bugs, where the number of unpersisted write bugs for correctness was 60 (highest), flushes to untouched memory (performance) bugs was 2 and fences with no commit (performance) bugs was 33. While there was no extra flush/fence performance bugs were detected. The results that AGAMOTTO identified is better than WITCHER bug detection tool as it only able to find 40 bugs compared to our validation tool discovered 60 new correctness PM bugs in the PMDK library. WITCHER also tested their PM applications (such as B-Tree,
RB-Tree, Hashmap-atomic, P-CLHT, Memcached and Redis including PMDK libraries \cite{witch}) over AGAMOTTO found same amount of performance bugs on their evaluation while level hashing was not tested by that time. So, our new level hashing program testing results are presented in the below table~\ref{table:tab}.

\begin{table}[ht]
    \centering
    \begin{tabular}{|c|l|l|l|l|l|l|}
    \hline
    \textbf{System}              & U-C &U-P & EP & Fl-P & Fe-P  & \textbf{Total} \\
    \hline
    WITCHER Level Hashing  & 17 & 11  & 12  & 0    & 0    & \textbf{40}    \\ \hline
    \textbf{Level Hashing (Ours)} & 60 & 0 & 0 & 2 & 3 & \textbf{65}  \\
    \hline
    \end{tabular}
    \caption{Tested Bug Results of NVM Level Hashing. Classes of bugs- U-C: Number of unpersisted write bugs (correctness); U-P: Number of unpersisted performance bugs; EP: Number of extra flush bugs (performance); Fl-P: Number of flushes to untouched memory (performance); Fe-P: Number of fences with nothing to commit (performance).}
    \label{table:tab}
\end{table}

\begin{figure}[htbp]
\centering
    \includegraphics[width=0.51\textwidth]{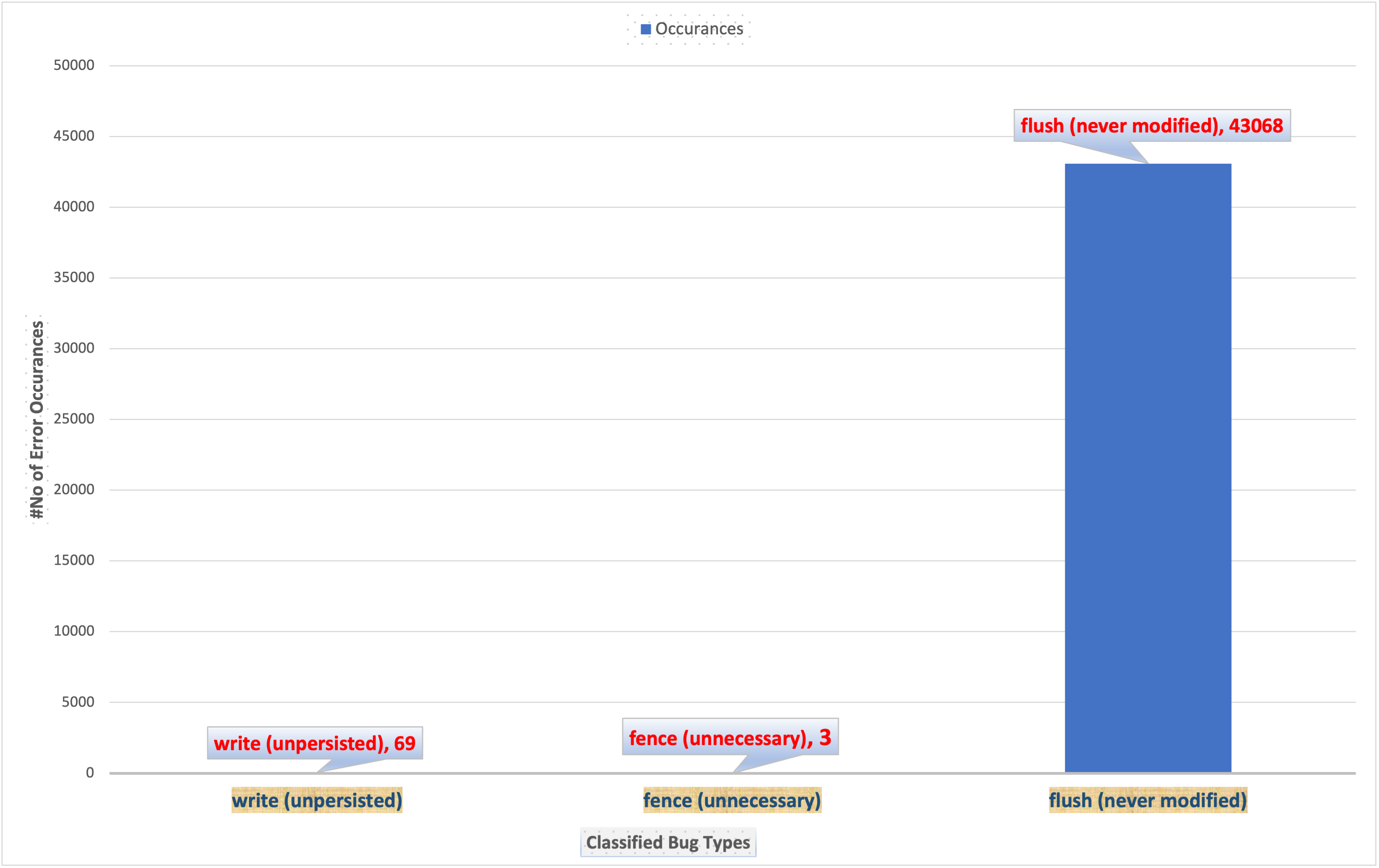}
    \caption{Classifying Bug Types With The Total Number Of Error Occurrences.}
    \label{fig:bug}
\end{figure}

Furthermore, most of the performance bug occurrences was for missing memory flush and which was $43,068$. Below figure~\ref{fig:bug} represents the classifying bug types with the total number of error or bug that occurrences during NVM path exploration search. Based on the reported instruction, it shows that flush that never modified have highest number of error occurrences ($43068$) compare to other unpersisted write ($69$) or unnecessary fence ($3$) bugs. Here, the write unpersisted bugs mean that there has been an update but it has never been flushed or fenced on the persistent main memory, which can indicate two different types of bugs commonly known as either performance (i.e. volatile data that does not need to be persistent) or correctness (i.e. the data should be made crash-consistent) bugs. Next, the unnecessary flush is being issued on data that already persistent in the flush where it never updated or modified. And, the fences are quite similar where its more likely unnecessary offenses. Based on our validation tool, the priori knowledge has a heuristic search to identify those bug oracles by defining some set of rules that indicate when a persistency bug has occurred in the systems. Although, it is quite hard to know how these rules perform on those PM specific hardware, is there a way to search those bug conditions? It might difficult to answer as each validation tools so far performed specific tasks based on the heuristic algorithm they used and to identify those bugs based on time period it takes so far. To leverage our idea, we will try to perform a study on Deep-Q learning algorithm to propose a way to improve our PM-Aware searching algorithm to have more efficient and robust state search exploration of program paths to detect those bugs based on its conditions.

\smallskip
Moreover, we will illustrate few of the bug listing below that has been observed and reported by our tool based on the execution testing results. Additionally, we will discuss two listings from the bug thread reports and study the possible root cause of such crash-consistency that occurred in our testing.
\smallskip

\begin{lstlisting}[language=Octave]
(gdb) list flush_clwb_nolog
90      /*
91       * flush_clwb_nolog -- flush the CPU cache, using clwb
92       */
93      static force_inline void
94      flush_clwb_nolog(const void *addr, size_t len)
95      {
96              uintptr_t uptr;
102             for (uptr = (uintptr_t)addr & ~(FLUSH_ALIGN - 1);
103                     uptr < (uintptr_t)addr + len; uptr += FLUSH_ALIGN) {
104                     pmem_clwb((char *)uptr); # Here the clwb iterating through like an arbitrary size which flushing the cache line
105             }
106     }
\end{lstlisting}
\captionof{lstlisting}{Bug Thread Found at line $104$ for $flush\_clwb\_nolog$.}

In the listing 2, we can see that the flush occurred at cache but still the $pmem\_clwb$ iterating through an arbitrary size of memory, which might be the reason for crash in this function calls. We used gdb to locate the bug for further analysis in this bug listing report. 

\begin{lstlisting}[language=Octave]
2240 /* opaque info lives at the beginning of mapped memory pool */
# Here hdr already persisted at line 2326,  here pool header is flushed without being entirely modified.
2241        struct pool_hdr *hdrp = rep->part[partidx].hdr;
.....
2235       /* store pool header */
# hdrp already persisted here
2236      util_persist_auto(rep->is_pmem, hdrp, sizeof(*hdrp)); 

\end{lstlisting}
\captionof{lstlisting}{Bug Thread Found at line $2236$ for $util\_header\_create$.}

At listing 3, we looked into the PMDK library at $src/common/set.c$ to locate the $hdrp$ pool header bug that detected by our validation tool. We can see that in this type of data structure the pool header seems flushed without any modification as the flushing here occurred in the pointer location in this whole header pool. But, we can observe that the header structure is larger than just the parts of it that are being modified. 

\subsection{Discussion and Lessons Learned} 
Since our project outcome is to identify the bugs based on the exploration path it took to detect those bugs and analyzes the generated report of bug classification types using our faithful testing tool. Over many proposed solutions, does these tools meet the expectations of their own to provide accurate results? This might be an open discussion to answer it, as our program analysis tool and the open-sourced workloads of level hashing might have different dependency that might be a reason of different results in our case. Also each bug detection tools are unique than each other, so, is there any trade-off parameters that can trigger to help with detection technique efficiently? For this, we might propose a universal heuristic algorithm to explore the bugs for all the testing tools but even we do so, it might have some biased results, so this might fail our expectation. Identifying PM bugs are quite challenging, does those heuristic search strategy helps to find all bugs based on bugs conditions? Yes, it does, even in our study all of the testing tools so far used their own searching algorithm and each of it generate satisfying experimental results. The key lessons that we learned from this project is that this depth study of PM bug detection help us to identify or locating bugs on PMDK library, then experimenting with some existing PM applications and reproduce the benchmark paper that we used for this piece of work.

\subsection{Suggestions And Further Improvements}
Based on our study, we found that detecting concurrent (i.e. not triggered in non-PM programs) inconsistent PM
bugs can be a primer focus to improve further in a key-value pair data structure. Also, our proposed methods might provide better state space heuristic search tree exploration path strategy to improve it further. Finally, study further study and experimenting few other existing tools to get more familiarity and use cases of these tools to detect real-world system bugs in general.

\section{Related Work}\label{section:relate}
\textbf{PM Programming and Testing Tools:} Over the years, a lot of research has been done to study and analyze the crash-consistency behavior on PM applications. Some of the prior works on PM programming was helpful to design for PM debugging and testing tools~\cite{xfd, pmtest, neal}, PM programming libraries and APIs~\cite{pmem, memo}.
And there has been many solutions approach has been designed and implemented to test and validate to improve the bug detection technique. In addition, Intel designed Yat~\cite{yat} to target specific crash-consistency bugs on PM file systems (PMFS)~\cite{fif} by computing all the possible ordering of its tracking instructions. Also, PMFuzz~\cite{pmfuzz} testing framework adopted fuzzing concepts to dynamically detect the NVM bugs by generating different input values on random test environment by fed WITCHER framework on it. Hippocrates~\cite{hipp} also an another automatic bug detection tool that can mitigate the NVM bugs by placing the flush and fences operations at the optimal memory address. Note that, the testing tool of AGAMOTTO is quite compatible artifact that we used to experiment our work of PM bugs analysis in this project.

\section{Conclusion and Future Work}\label{section:con}
In this paper, we experimented NVM level hashing testing program on our AGAMOTTO validation tool to detect the NVM bugs on PMDK library. To provide persisted key-value pair on our persistent memory to validate the performance and correctness bugs by an automatic inspection checker. In our adopted automatic symbolic execution bug detector tool, we able to detect 65 new PM bugs on NVM level hashing testing program with a minor modification from the WITCHER source code to test on our experimental environment. This extensive test results showed that the AGAMOTTO able to detect more bugs compared to the WITCHER framework, which able to detect only 40 bugs in the level hashing testing program and all incurs no false positives reported bugs. As a part of future work, this paper still aims to investigate further to propose a fruitful search selection algorithm based on our proposed methods that discussed earlier.

\section*{Acknowledgment}
We sincerely thanks Ian Neal, a PhD candidate studying Computer Science at the University of Michigan for his consistent support on explaining AGAMOTTO paper, as this whole project work was based on his paper. The author declares that there is no conflict of interest.

\vspace{12pt}

\end{document}